# FUNDAMENTAL VIEW ON THE CALCULATION OF INTERNAL PARTITION FUNCTIONS USING OCCUPATIONAL PROBABILITIES

Mofreh R. Zaghloul

Department of Physics, College of Sciences, United Arab Emirates University,

P.O.Box 17551, Al-Ain, UAE.

**ABSTRACT**

From first principles, the author gathers a few general rules that need to be abided by in the calculation of the internal partition functions (IPFs) of individual molecules. These rules are violated in many schemes in the literature where occupational probabilities are used including those using the Planck-Larkin partition function (PLPF) within the chemical picture. Considering these rules is useful from conceptual and practical points of view. A solution is introduced to assure fulfilling the above mentioned rules when using occupational probabilities. Sample calculations are performed showing quantitative significance of inaccuracies caused by dishonoring such rules.

# I- INTRODUCTION

The problem of establishing finite and statistical-mechanically consistent internal partition functions in nonideal plasma systems, within the chemical picture, has been recently revisited and revised [1-4]. In these studies, it was shown that any separable configurational component of the free energy is equivalent to a common factor (independent of the individual excited states) multiplied by all individual terms in the sum over states, which can be factored out again with no implications on the divergence problem of the IPF. It was therefore concluded that the use of a separable configurational component of the free energy function cannot exclusively lead to the establishment of finite statistical-mechanically consistent electronic partition functions unless an abrupt cutoff scheme is adopted. In that context it was shown that formulae and formulations similar to those by Fermi [5,6], Hummer and Mihalas [7], and Potekhin [8] are either inaccurate or statistical-mechanically inconsistent. Further, a remedy for the problem was proposed in terms of the solution of the inverse problem in which, and based on physical bases, an occupational probability is prescribed in advance to assure a smooth truncation of the IPF and a manifestation of the phenomenon of pressure ionization. The resulting IPF can be used, therefore, to calculate the occupation numbers and to find the corresponding set of modified thermodynamic properties. The solution is a materialization of the fundamental statistical thermodynamic fact indicating that the equilibrium behavior of matter can be completely understood from the partition function.

With occupational probabilities or bound-state weighting factors, the internal electronic partition function can be written as

$$Q_{int}^{elec}(V,T,\{N\}) = \sum_{i=0}^{\infty} g_i\, w_i(V,T,\{N\})\, exp\left(\frac{-\varepsilon_i}{K_B T}\right) \qquad (1)$$

where $V$ is the volume of the system, $T$ is the absolute temperature, $K_B$ is the Boltzmann constant, $\{N\}$ are the occupational numbers of particles, $g_i$ and $\varepsilon_i$ are the statistical weight and the unperturbed excitation energy above the ground state for the $i$th excited state, respectively. The weighting factor $w_i(V,T,\{N\})$ is the corresponding state-dependent occupational probability of that level. According to Hummer and Mihalas [7] the occupational probability, $w_i(V,T,\{N\})$, is presumed to decrease *continuously* and *monotonically* as the strength of the relevant interactions increases in order to produce a physically reasonable continuous transition between bound and free states. Further, the occupational probability $w_i(V,T,\{N\})$ should drop strongly to zero as the binding energy of a level below the unperturbed continuum goes to zero in order to provide natural and smooth truncation of the internal partition function. In the *step-cutoff* scheme, commonly used to establish finite bound-state partition functions, the occupational probability is simply represented by a unit step function where the weighting factor $w_i(V,T,\{N\})$ is represented as

$$\begin{cases} w_i = 1. & \text{for } \varepsilon_i < I - \Delta I \\ w_i = 0. & \text{for } \varepsilon_i \geq I - \Delta I \end{cases} \qquad (2)$$

where $\Delta I$ refers to a term traditionally known as *lowering of ionization energy*. On the other hand, different continuous forms of occupational probabilities have been proposed and implemented in the literature (see for example references [5-11]). A common factor among all of these continuous forms is that $w_i(V,T,\{N\})$ decreases *continuously* and *monotonically* as particle densities increase.

The Planck-Larkin partition function (PLPF), regularly encountered in the framework of the "*physical picture*" is convergent without additional cut-offs. This fact led many authors in the literature to propose and use it in the calculations of the equation of state of plasma systems within the "*chemical picture*" (see for example Refs. [12-15]). Nevertheless, such implementation of the PLPF in the chemical model was controversial, mainly for being in contradiction with observations (see for example Refs. [16-18]). According to Däppen, et al [18], "experimental evidence shows that it is inconsistent to use the Planck-Larkin partition function as the internal partition function in simple models of reacting gases (i.e., the "chemical picture")". It is useful, for the sake of the analysis given herein, to realize that implementing the PLPF is equivalent, in every respect, to using a temperature-dependent occupational probability, $w_{i,elec}^{PLPF}(T)$, in Eq. (1) where

$$
\begin{aligned}
Q_{int,elec}^{PLPF}(T) &= e^{E_0/K_BT} \sum_{i=0}^{\infty} g_i \left[ e^{-E_i/K_BT} - 1 + \frac{E_i}{K_BT} \right] \\
&= \sum_{i=0}^{\infty} g_i \left[ 1 - e^{E_i/K_BT}\left(1 - \frac{E_i}{K_BT}\right) \right] e^{-\varepsilon_i/K_BT} \\
&= \sum_{i=0}^{\infty} g_i \left[ 1 - e^{(\varepsilon_i-I)/K_BT}\left(1 - \frac{(\varepsilon_i-I)}{K_BT}\right) \right] e^{-\varepsilon_i/K_BT} \\
&= \sum_{i=0}^{\infty} g_i \, w_{i,elec}^{PLPF}(T) e^{-\varepsilon_i/K_BT}
\end{aligned}
\qquad (3)
$$

In the above equation, $E_i$ is the energy of the *i*th energy level with respect to the continuum, $\varepsilon_i$ is the energy of the same level but relative to the ground state, $g_i$ is its statistical weight, and $I=-E_0$ is the ionization energy of the ion under consideration.

In a response to Rouse's criticism [16], Ebeling et al in Ref. [19] recommended using the discrete energy states of the Bethe-Salpeter equation (BSE) in the PLPF, for quantum statistical Coulomb systems with bound states, in order for PLPF to be density-dependent as well. Yet, the minimization of free energy, in such a case, would embody the derivatives of the partition function and the density-dependent discrete energies with respect to occupation numbers (see relevant discussion in Refs. [1-3]).

## II- FUNDAMENTALS

The present section presents a recollection of some relevant fundamentals of statistical thermodynamics. Although the information presented herein is very basic, the comprehension of these fundamentals is essential to derive a set of general requirements that need to be fulfilled in the calculation of the IPF and to explain conceptual errors found in the literature.

The Helmholtz free energy function for a canonical ensemble can be expressed in terms of the canonical partition function of the system, $Z_{tot,sys}$ as

$$F = -KT \ln Z_{tot,sys} \tag{4}$$

where $Z_{tot,sys}$ is the total canonical partition function of the *system* or *assembly* which can be expressed as

$$\begin{aligned} Z_{tot,sys} &= \sum_{states} e^{-E_{i,sys}/K_B T} \\ &= e^{-E_{0,sys}/K_B T} \sum_{states} e^{-\varepsilon_{i,sys}/K_B T} \\ &= e^{-E_{0,sys}/K_B T} \overline{Z}_{tot,sys} \end{aligned} \tag{5}$$

where $E_{i,sys}$ is the energy of the *i*th quantum state of the whole system or assembly, $\varepsilon_{i,sys}=(E_{i,sys}-E_{0,sys})$ is the energy of the same quantum state but relative to the lowest or ground state energy and $\overline{Z}_{tot,sys}$ is the scaled bulk-state partition function calculated using energy states relative to the ground state of the system. Taking Eq. (5) into consideration, equation (4) can also be rewritten as

$$F - E_{0,sys} = -KT \ln \overline{Z}_{tot,sys} \tag{6}$$

The system or assembly may consist of any number of molecules in any sort of interaction with one another. Thus the assembly may be solid, liquid, or gas.

For an assembly of non-interacting molecules, statistical theory makes possible the expression of the bulk-state functions in terms of the relevant properties of the individual molecules. For a non-interacting assembly consisting of several kinds of molecules, the bulk state partition function may be factored into a product of co-factors each corresponding to the particles of one kind, i.e.

$$\begin{aligned} Z_{tot,sys} &= \prod_s \frac{Q_{tot,s}^{N_s}}{N_s!} \\ &= \frac{Q_{tot,A}^{N_A}}{N_A!} \frac{Q_{tot,B}^{N_B}}{N_B!} \cdots \frac{Q_{tot,e}^{N_e}}{N_e!} \prod_{j,\zeta} \frac{Q_{tot,j,\zeta}^{N_{j,\zeta}}}{N_{j,\zeta}!} \end{aligned} \tag{7}$$

where $Q_{tot,s}$ is the total partition function of the particle of kind *s* and $N_s$ is the number of these particles in the assembly with the index, *s*, running over all kinds of particles. In the second line of the above equation the subscripts A, B,… refer to polyatomic molecules, the subscript *e* refers to free electrons while the subscripts *j* and $\zeta$ refer in order to the chemical element and the multiplicity of the atomic ion. The word "*molecule*" is generally used to refer to any particle whether polyatomic, atomic/ionic or free electron. Nevertheless, polyatomic molecules are excluded from the following discussion as the main focus of the present work is the bound state *electronic* partition function.

The total *molecular* partition function is expressed by equations of a form similar to Eq. (5) where

$$Q_{tot} = \sum_{states\ k} e^{-E_{k,tot}/K_B T} \tag{8}$$

Here $E_{k,tot}$ is the $k$th quantum state of the *total* energy of the molecule under consideration and the summation is carried out over all possible states of the molecule. If the energy levels are degenerate, then each contribution enters the partition function as an independent component, so that the number of identical components of each level is equal to the statistical weight, $g$, of the level. Hence,

$$Q_{tot} = \sum_{states\, k} e^{-E_{k,tot}/K_B T} = \sum_{levels\, r} g_r\, e^{-E_{r,tot}/K_B T} \qquad (9)$$

It has to be remembered that the energy of a composite molecule (atom or ion) is the sum of two *distinct* parts; translational part, $E_{trans}$, and internal part, $E_{int}$, which is due to non-translational causes such as electronic excitation. However, *the possession of a particular amount of internal energy is entirely independent of the magnitude of the translational energy, and vice versa.* Therefore, for the molecular partition function, $Q_{tot}$, one can write

$$\begin{aligned}Q_{tot} &= \sum_{all\, states} e^{-(E_{trans}+E_{int})/K_B T} = \sum_{trans.\, states} e^{-E_{trans}/K_B T} \times \sum_{int.\, states} e^{-E_{int}/K_B T} \\ &= \sum_{trans.\, levels} g_{trans}\, e^{-E_{trans}/K_B T} \times \sum_{int.\, levels} g_{int}\, e^{-E_{int}/K_B T} \\ &= Q_{trans}\, Q_{int}\end{aligned} \qquad (10)$$

where the summation $\sum_{all\, states}$ is extended over all the translational and internal states while the sums $\sum_{trans.\, states}$ and $\sum_{int.\, states}$ are extended over all translational states and over all internal states, respectively. Equation (10) clearly indicates that for a non-interacting assembly, the total molecular partition function is factorizable. Applying the factorizability of the *total molecular* partition function to Eq. (9) and substituting back into Eq. (7) and Eq. (4) one straightforwardly arrives at

$$\begin{aligned}F &= -K_B T \ln Z_{tot,sys} \\ &= -K_B T \sum_s N_s \ln\left(\frac{Q_{trans,s}}{N_s!}\right) - K_B T \sum_s N_s \ln Q_{int,s}\end{aligned} \qquad (11)$$

As it can be seen from Eq. (11), the factorizability of the total molecular partition function shown in Eq. (10) is equivalent to separability of the free energy function and vise versa. Accordingly, the tradition of adding a configurational component to the free energy function in order to account for particle interactions is equivalent, in every respect, to multiplying the partition function by a configurational factor. The free energy function in this case becomes

$$\begin{aligned}F &= -K_B T \sum_s N_s \ln\left(\frac{Q_{trans,s}}{N_s!}\right) - K_B T \sum_s N_s \ln Q_{int,s} + F_{conf} \\ &= -K_B T \sum_s N_s \ln\left(e^{\frac{-F_{conf}}{\sum_s N_s K_B T}} \frac{Q_{trans,s}}{N_s!}\right) - K_B T \sum_s N_s \ln Q_{int,s}\end{aligned} \qquad (12)$$

In Eq. (12), the configurational factor is collected with the translational partition function for no reason but to show that it has no influence on the calculation of the internal partition function.

## III- GENERAL REQUIREMENTS

From the above basic review one can gather a few general requirements that need to be fulfilled in the calculation of the internal partition function IPF:

i. *It is not allowed for the internal partition function of any individual particle to be ≤zero*. This can be easily recognized since the translational factor is always positive and the arguments of the logarithms in Eqs. (3,9,10) should always by positive.

ii. *Within the step cut-off theory of the internal electronic partition function, the lowering of ionization energies should not exceed in magnitude the ionization energy itself*. This requirement is in compliance with the above stated one; otherwise the internal partition function would be zero, which is forbidden according to the first requirement. It has to be noted that the majority of widely used models for the lowering of ionization energies do not satisfy this requirement at very high densities.

iii. *The internal partition function should not, in any case, be less than unity*. This is a more important and a more restrictive requirement. It can be easily recognized by recalling that *the possession of a particular amount of internal energy is entirely independent of the magnitude of the translational energy, and vice versa*. Logically, having an internal structure and internal energy states (electronic for example) should increase the total number of all states according to Eq. (10) ($Q_{tot} = \sum_{all\ states} e^{-(E_{trans}+E_{int})/K_BT}$) where the total number of all states equals the number of translational states times the number of internal states. Moreover, not only the total number of all states will be increased by having internal energy states but also for each new state the energy becomes ($E_{trans}+E_{int}$) and since $E_{int}$ for bound electronic states is negative, then the factor $e^{-E_{int}/K_BT}$ is larger than unity with the result that the internal electronic factor of the partition function must be ≥1. It has to be remembered that the absence of any internal energy states means that the internal factor of the partition function is unity. This important finding is terribly violated when occupational probabilities are used and in particular when the PLPF is used for the case of plasmas at very high temperatures as we show below.

## IV- VIOLATIONS OF THE FUNDAMENTAL REQUIREMENTS AND THEIR CONSEQUENCES

In view of requirements *i* and *iii* explained above, it appears that using the Planck-Larkin partition function or Planck-Larkin occupational probability violates both requirements as it effectively approaches zero at very high temperature. To show this, one can easily expand the exponential inside the sum in the first line in Eq. (3) to get

$$Q_{int,elec}^{PLPF}(T) = e^{E_0/K_BT} \sum_{i=0}^{\infty} g_i \left[ \left(1 - \frac{E_i}{K_BT} + \frac{E_i^2}{2!(K_BT)^2} - \frac{E_i^3}{3!(K_BT)^3} + ... \right) - 1 + \frac{E_i}{K_BT} \right]$$

$$= e^{E_0/K_BT} \sum_{i=0}^{\infty} g_i \left[ \frac{E_i^2}{2!(K_BT)^2} - \frac{E_i^3}{3!(K_BT)^3} + ... \right]$$

(13)

Since the bound states energies are bound by the difference between the continuum and the lowest or ground energy, one concludes that as $T \to \infty$, $Q_{int,elec}^{PLPF} \to 0$. Accordingly, using the Planck-Larkin occupational probability, within the chemical model is conceptually incorrect as it violates the above mentioned requirements. Figure 1 shows isotherms of the contribution of each term in the first 10 terms in the electronic IPF of hydrogen for the Planck-Larkin occupational probability (thick lines) in comparison to the case of a unity occupational probability (thin lines). As it can be seen from the figure, although the contribution of the first term for the case of $w_i = 1$ is always greater than unity and equals to the statistical weight of the ground state ($g_0=2$), the contribution of the same term when using the Planck-Larkin occupational probability decreases with temperature to values less than unity and goes to zero as the temperature goes to infinity.

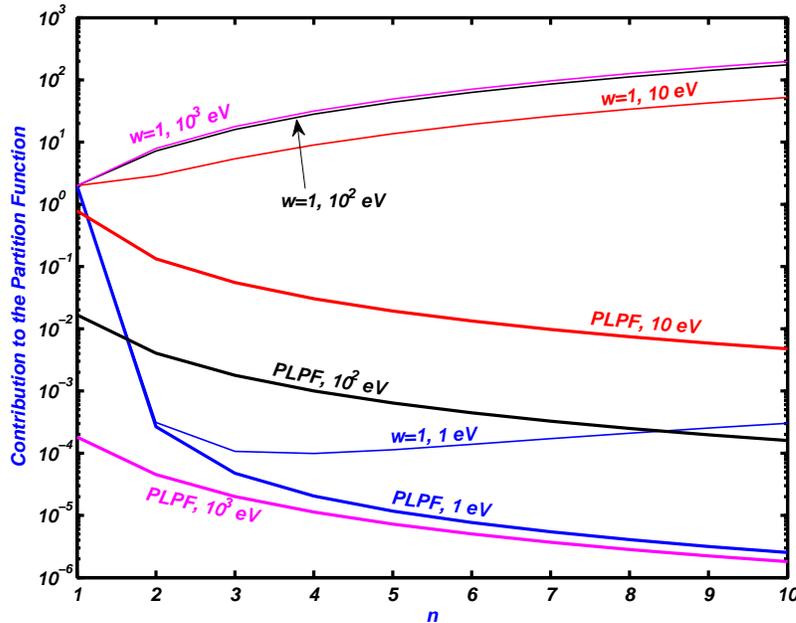

Figure 1. Isotherms of the contribution of each term in the first 10 terms in the electronic IPF of hydrogen for the Planck-Larkin occupational probability (thick lines) in comparison to the case of a unity occupational probability (thin lines)

Figures 2-a and 2-b show isotherms of the sum of the terms as a function of the number of terms considered in the calculation of the electronic IPF of hydrogen for both of the case of Planck-Larkin occupational probability and the case of $w_i = 1$, respectively. One can clearly see the convergent nature of the PLPF, however, its values decrease monotonically with the temperature and become not only *less than unity* but also they approach zero as the temperature goes to infinity.

On the other hand, continuous and state-dependent forms of the occupational probabilities, similar to those in Refs [5-11], are all monotonically decreasing with density and approach zero at high densities. This happens even with the occupational probability of the ground state. As a result, the calculated electronic partition function at very high densities will not only drop to values lower than unity, in violation of requirement *iii*, but will also approach zero at very high densities, in violation to requirement *i* too.

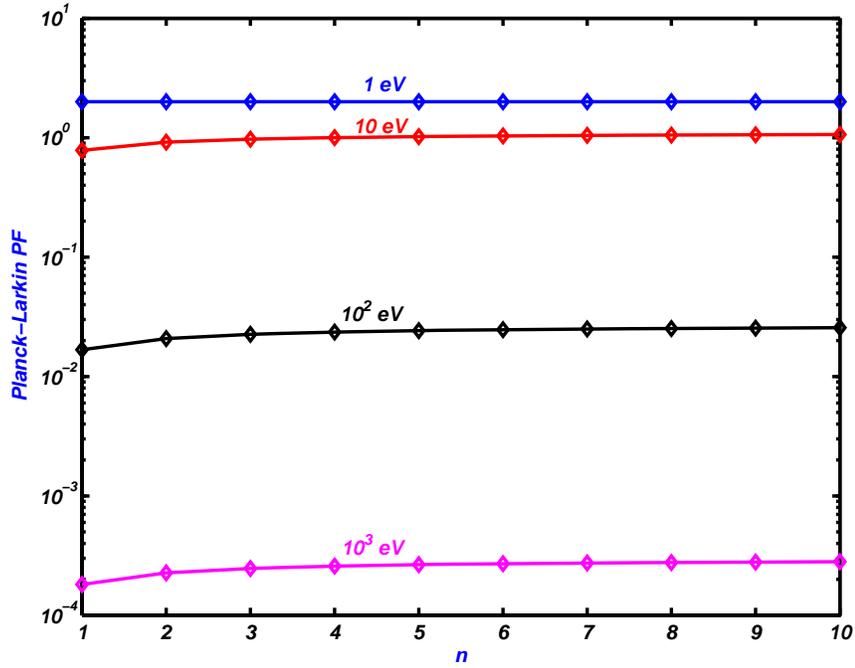

Figure 2-a. The sum of terms in the electronic IPF of hydrogen as a function of the number of terms using Planck-Larkin occupational probability

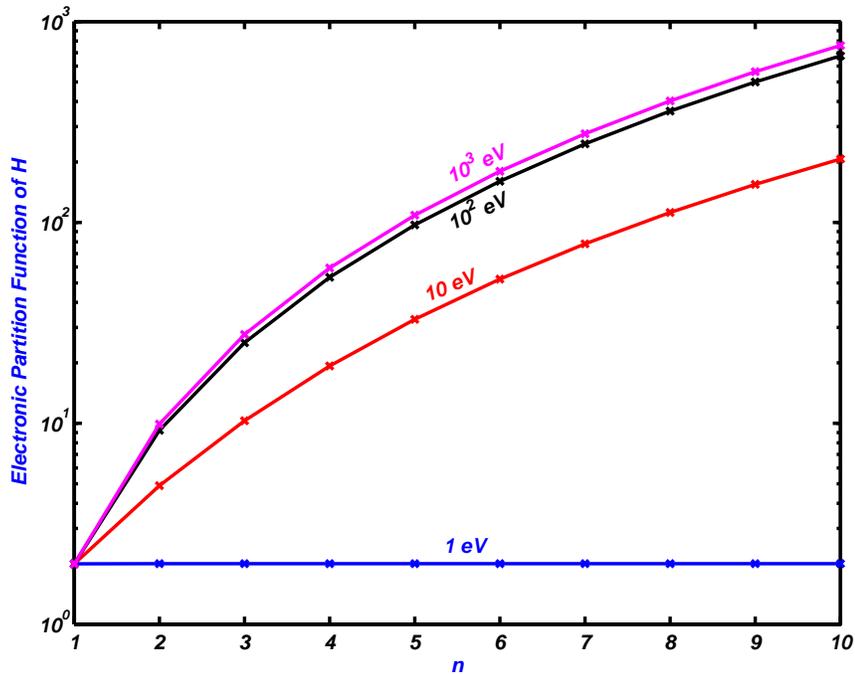

Figure 2-b. The sum of terms in the electronic IPF of hydrogen as a function of the number of terms using $w_i = 1$

It may be important to advise here that the above mentioned violations exist in a massive number of publications in the literature and in models and computations developed by distinguished research groups. A pressing question, however, is; what is the impact of these violations on the calculation of ionization equilibrium of partially ionized plasma, which is the first step in the calculation of the equation of state, thermodynamic properties and optical characteristics?

To give a rough answer to the question raised above, we consider a simple case of ideal, partially ionized and non-degenerate plasma generated from a single chemical element. Using the temperature-dependent PLPF, the conditions for minimization of the free energy lead to a system of equations of the form [1-2]

$$K_B T \ln\left(\frac{N_\zeta}{N_{\zeta+1}} \frac{2V Q_{int\zeta+1}(T)}{N_e \Lambda_e^3 Q_{int\zeta}(T)}\right) - I_\zeta = 0, \qquad \zeta = 0, 1, \ldots, Z \qquad (14)$$

where $\Lambda_e = h/\sqrt{2\pi m_e K_B T}$ is the average thermal de Broglie wave length of electrons, $I_\zeta$ is the ionization energy of the ion $\zeta$, and $Z$ is the atomic number of the chemical element. This system of equations needs to be solved subjected to the constraints of electroneutrality and conservation of nuclei. As it can be seen from equations (14), the electronic IPF appears in both of the numerator and denominator of the argument of the logarithm which means that the consequences of violating the above mentioned requirements would not be prominent and can be alleviated in some cases. Nevertheless, the consequences of violating the above requirements (particularly, requirement *iii*) should become obvious for a case like high-temperature hydrogen plasma where the electronic IPF for protons is constant (unity) independent of the state of the system. This should also be the case for plasmas with high ionization states where bare nuclei play an important role in the calculation of ionization equilibrium. For such cases, the consequences of violating requirement *iii* in the denominator of the argument of the logarithm in the last equation of Eqs (14) would not be alleviated by the constant (unity) state-independent IPF of the nucleus in the numerator of the argument of the logarithm.

Using number densities and applying the constraints of electroneutrality and conservation of nuclei, the problem of determining the ionization equilibrium for the case of hydrogen is reduced to solving the following equation,

$$\left(\frac{(1-\zeta_{av})}{\zeta_{av}^2} \frac{2(2\pi m_e K_B)^{3/2} (Q_p = 1) T^{3/2}}{n h^3 Q_{H,PLPF}}\right) = \exp\left(\frac{I_H}{K_B T}\right) \qquad (15)$$

where $n$ is the number density of heavy particles, $\zeta_{av} = n_p/n$ is the average ionization state with $n_p$ representing the number density of protons and $I_H$ is the ionization energy of neutral hydrogen. For high temperatures where $K_B T \gg I_H$, the right hand side of the above equation approaches unity and one gets

$$\frac{\zeta_{av}^2}{(1-\zeta_{av})} \propto \frac{T^{3/2}}{Q_{H,PLPF}} \qquad (16)$$

Now, since the PLPF decreases monotonically and gets less than unity as $T$ increases, the degree of ionization predicted by using the PLPF at high temperatures will be overestimated as it is clear from (16), which leads to depletion of neutral and excited states. Although this rough answer to the above raised question has been obtained for the simple case of hydrogen plasma using Planck-Larkin occupational probability, it is indicative of the consequences for other real situations when using occupational probabilities that violate the above discussed requirements, particularly requirement *iii*.

At this point, it may be enlightening to recall that a vanishing IPF, as expected by the PLPF at very high temperatures or by a continuous state-dependent occupation probability SDOP at high densities, does not mean and should not be interpreted as zero occupation of the specific composite particle under consideration, but rather it indicates that the entity of such a composite particle does not exist at all destroying the problem foundation within the framework of the chemical model.

## V- A REMEDY AND SAMPLE CALCULATIONS

In view of the above discussed requirements, a remedy that guarantees the fulfillment of all of the requirements in section-III, when using discrete or continuous occupational probabilities can be easily obtained by writing the electronic IPF in the form

$$Q_{int}^{elec}(V,T,\{N\}) = [1 - w_0(V,T,\{N\})] + \sum_{i=0}^{\infty} g_i w_i e^{-\varepsilon_i/K_B T} \qquad (17)$$

where the energy $\varepsilon_i$ is measured from the ground state and a term $[1-w_0(V,T,\{N\})]$ is added to the traditional expression of the IPF with occupational probability where $w_0$ is the value of the occupational probability for the ground state. The contribution of the ground state alone can, therefore, be written as

$$Q_0(V,T,\{N\}) = [1 + (g_0 - 1) w_0(V,T,\{N\})] \qquad (18)$$

For extreme cases where values of the expression for $w_i$ (including $w_0$) approach zero, the value of the IPF will converge to unity, in accordance to the above discussed requirements. This simply indicates that even for extreme cases where all internal states (including degenerates of the ground level) are strongly perturbed or removed, the internal factor of the partition function will be unity as mandated by the above discussed requirements and as physically expected for a simple non-degenerate particle. On the other extreme where the value of $w_0$ is unity, the contribution of the ground state will, therefore, be equal to its statistical weight and the expression in Eq. (17) simply reduces to the traditional commonly used expression. The value of the IPF in this case will be larger than or equal to the statistical weight of the ground state as expected. The above introduced remedy (or form) should be used for the IPF when occupational probabilities are used to assure satisfaction of the above-mentioned requirements.

Sample calculations of ionization equilibrium are performed to show the effect of implementing the remedy introduced above on the calculation of the plasma composition. Figures 3 and 4 show the fractions of protons, $\zeta_{av}$, and of neutral hydrogen atoms, $\alpha_0$, calculated for a 20 eV and 1.5 eV hydrogen plasma over a wide range of densities. The fractions of protons and neutral hydrogen atoms are defined as $\zeta_{av}=n_p/n$, and $\alpha_0= n_H/n$ where $n_p$ is the number density of protons and $n_H$ is the number density of neutral hydrogen atoms while $n=n_p+n_H$ is the total number density of heavy particles or nuclei in the system. The figures show four groups of curves representing: (a) results using the traditional temperature-dependent PLPF, (b) results using the temperature-dependent modified Planck-Larkin partition function, MPLPF, where the remedy or modification introduced above is implemented, (c) results using a state-dependent occupational probability, SDOP (see Ref. [10] for example), and finally (d) results using a modified state-dependent occupational probability, MSDOP, where the modification or the remedy introduced above is implemented to the occupational probability used in (c). Partial degeneracy of free electrons and Coulomb's nonideality corrections have been taken into account in these calculations. For better description of these methods and their results, we write the equation used for the calculation of the ionization equilibrium for the case of hydrogen in its full form

$$\frac{\zeta_{av}^2}{1-\zeta_{av}} = C \times \frac{T^{3/2}}{n\, Q_H(n,T)} \times \exp\left(\frac{-\left[I_H - \Delta I_H^{F_C} - \Delta I_H^{dgc} - \Delta I_H^{Q^{int}}\right]}{K_B T}\right), \quad (19)$$

where $C = 2\frac{(2\pi m_e K_B)^{3/2} Q_p(=1)}{h^3}$ is a constant independent of density or temperature.

In the above equation $\Delta I_H^{F_C}$, $\Delta I_H^{dgc}$, and $\Delta I_H^{Q^{int}}$ are in order; the magnitudes of depression of ionization potential due to Coulombic nonideality corrections, virtual upshifting of ionization potential due to partial degeneracy of the electron gas, and the correction to the ionization potential due to the dependence of the IPF on density or pressure. Detailed expressions used for these terms in the present calculations can be found in Refs. [1,2]. For cases (a) and (b) stated above, $Q_H = Q_H(T)$, a function of $T$ only and therefore the last correction to the ionization potential, $\Delta I_H^{Q^{int}}$, vanishes. However, for cases (c) and (d), $Q_H = Q_H(n,T)$, a function of both $n$ and $T$ and the last correction to the ionization potential, $\Delta I_H^{Q^{int}}$, will stick around. As the temperature increases, the magnitude of the argument of the exponential factor in Eq. (19) decreases and the sensitivity of the factor to dependence on density fades out. The ionization fraction roughly follows the pre-exponential factor $\frac{\zeta_{av}^2}{1-\zeta_{av}} \approx C \times \frac{T^{3/2}}{n\, Q_H(n,T)}$ in this case. On the other hand as the temperature decreases, the magnitude of the argument of the exponential factor increases and the sensitivity of the factor to the dependence on density increases as well and it plays a major role in determining the degree of ionization.

     For the 20 eV hydrogen plasma, as it can be clearly seen from Fig. 3, the average ionization state (proton fraction) predicted by using the traditional PLPF is recognizably higher than that expected by using the modified PLPF, in which the above introduced remedy is implemented. Generally, the deviation between the suggested modified PLPF and the original PLPF is expected to appear significantly at high temperatures where the PLPF violates the third requirement. However, the *relative* magnitude of this deviation will also depend on the degree of ionization. Accordingly, the deviation will be relatively large at high temperatures (where the PLPF violates requirement *iii*) and high densities where recombination and degeneracy become important and the plasma is partially ionized. The higher ionization is equivalent to depopulation of the neutral atoms and consequently depopulation of the excited states. The same remarks are also valid when comparing calculations using the state-dependent occupational probabilities, SDOP with the traditional expression of the IPF and its modified version MSDOP given by Eq. (17). However, it has to be noted that for such a relatively high temperature isotherm as

explained above, the sensitivity of the exponential factor in Eq. (19) to the dependence on density weakens and the dependence of the degree of ionization on density is mainly reflected in the pre-exponential factor that is $1/nQ_H$. For both of PLPF and MPLPF, $Q_H$ does not depend on $n$ and as a result the degree of ionization predicted by these two methods decreases as with $n$ as shown in the figure. The modified SDOP does not allow unlimited sharp decrease of $Q_H$ with density and the degree of ionization predicted by this method decreases with $n$ as well. On the other hand the SDOP method which does not satisfy the above mentioned requirements, suffers unlimited sharp decrease of the $Q_H$ with $n$ which causes a remarkable increase in the degree of ionization giving rise to what is known as pressure ionization. This remarkable change of the behavior of the ionization equilibrium at high densities when using the MSDOP compared to the traditional SDOP could significantly affect our expectations of the equation of state, thermodynamic and optical properties in this region of the domain which will be investigated in following studies. It has to be noted, as a final remark on the computational results in figure 3, that the computational results from the MSDOP and MPLPF are relatively close in behavior over most of the computational domain.

The same calculations are presented in Fig. 4 but for a relatively low temperature (1.5 eV). As it can be seen from the figure, the predictions using all different schemes for the calculation of IPF, stated above, are mostly identical at relatively low densities $<10^{22}$ cm$^{-3}$. Predictions from the PLPF and MPLPF continue to be identical over the whole domain of densities as expected, for such a relatively low temperature. The close behavior of the predictions from MPLPF and MSDOP, over the entire density domain, continues for such a low temperature isotherm. The remarkable change of the behavior of the ionization equilibrium at high densities when using the MSDOP compared to the traditional SDOP continues at such a low temperature as well. The so-called pressure ionization (lowering of ionization energies) appears clearly at densities $>10^{22}$ cm$^{-3}$ before it gets suppressed by electron degeneracy (virtual upshifting of ionization energies, see for example Ref. [20]) at much higher densities except for the case of SDOP due to a violation of the requirements in section III as the IPF of neutral hydrogen gets less than unity and suffers unbounded decrease for such high densities. All methods show pressure ionization at such a low temperature due the enhancement of the sensitivity of the exponential factor in Eq. (19) to the dependence on density where the factor prevails causing an increase in ionization till it gets suppressed by electron degeneracy.

It has to be noted that the Planck-Larkin occupational probability effectively limits the number of excited states to those having binding energy $\geq K_BT$. In the framework of in which unperturbed energy levels of an isolated atom are used, this effective binding energy of the last level included in the PLPF, $\sim K_BT$, is in no way related to the plasma density in general (or to the lowering of ionization energy in particular) although it is the plasma density (or the lowering of ionization energy) that must define the binding energy at which an electron may be considered free (see relevant discussion in Ref. [22]). Finally, it may be useful from the practical point of view to recall that the number of observed lines in the solar photosphere is reportedly much larger than predicted by the PLPF as the PLPF effectively limits the number of excited states to those having binding $\geq KT$ (see for example Refs. [16-18]). Nevertheless, Rogers (1986) explained this fact by allowing resonances that are not counted in the partition function but could be seen in optical spectra. Accordingly, the controversy about the PLPF may not be fully resolved with optical experiments and measuring transport and thermodynamic properties is therefore desirable.

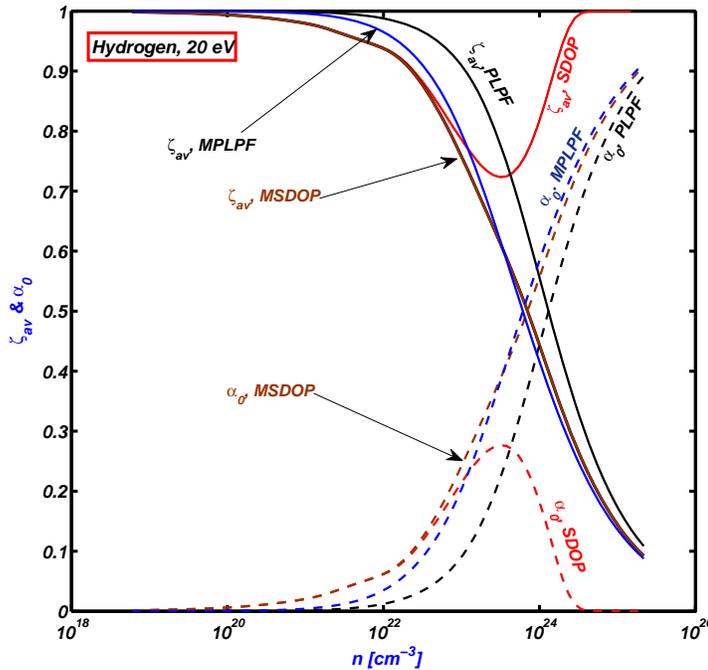

Figure 3. Fractions of protons, $\zeta_{av}$, and of neutrals atoms, $\alpha_0$, calculated for a 20 eV hydrogen plasma over a wide range of densities

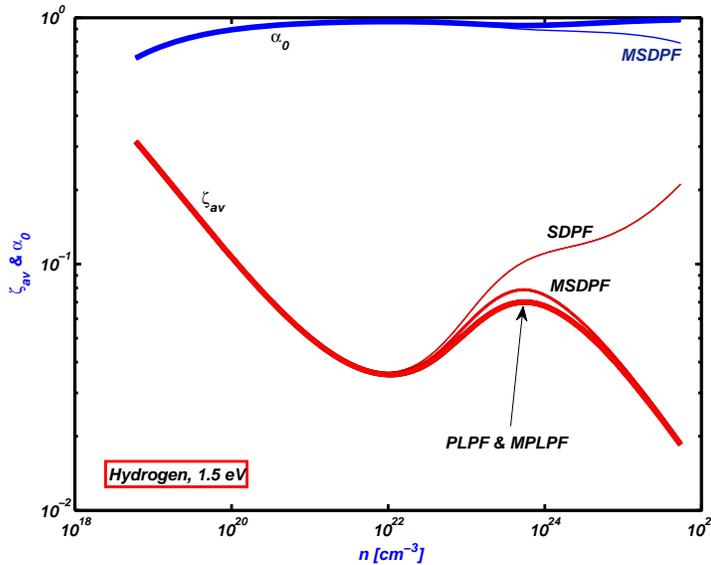

Figure 4. Fractions of protons, $\zeta_{av}$, and of neutrals atoms, $\alpha_0$, calculated for a 1.5 eV hydrogen plasma over a wide range of densities

## VI- CONCLUSIONS

A few general requirements that need to be fulfilled in the calculation of the internal partition functions of composite particles have been derived from first principles. It has been shown that many schemes used in the literature, for the calculation of the IPFs violate these rules. In particular, it has been shown that the widely used Planck-Larkin partition function violates these rules, when employed within the chemical picture, which adds a new critique to using PLPF within the chemical model. The consequences of these violations have been assessed for the case of hydrogen plasma and were found to be significant. A remedy for the problem is introduced and sample calculations are performed showing quantitative significance of inaccuracies caused by such violations and conceptual errors. The introduced remedy could help improving the behavior of the tempting formula of the Planck-Larkin partition function.